\documentclass[fleqn,usenatbib]{mnras}

\usepackage{newtxtext,newtxmath}
\usepackage[T1]{fontenc}
\usepackage{ae,aecompl}

\usepackage{graphicx}	% Including figure files
\usepackage{amsmath}	% Advanced maths commands
\usepackage{amssymb}	% Extra maths symbols
\usepackage{threeparttable} %for table footnotes - EXTRA
\usepackage{booktabs,siunitx} %for horizontal column dividers 
\usepackage{array,tabularx,ragged2e}
\title[Resolving the retired A-star mass controversy]{Asteroseismic masses of four evolved planet-hosting stars using SONG and TESS: resolving the retired A-star mass controversy}

\author[Malla et al.]{
Sai Prathyusha Malla$^{1}$\thanks{E-mail: s.malla@student.unsw.edu.au (UNSW)},
Dennis Stello$^{1, 2, 3}$,
Daniel Huber$^{4}$,
Benjamin T. Montet$^{1}$,
\newauthor
Timothy R. Bedding$^{2,3}$,
Mads Fredslund Andersen$^{3}$, 
Frank Grundahl$^{3}$,
\newauthor
Jens Jessen-Hansen$^{3}$,
Daniel R. Hey$^{2,3}$,
Pere L. Palle$^{5, 6}$,
Licai Deng$^{7}$,
Chunguang Zhang$^{7}$,
\newauthor
Xiaodian Chen$^{7}$,
James Lloyd$^{8}$,
Victoria Antoci$^{3,9}$
\\
$^{1}$School of Physics, The University of New South Wales, Sydney NSW 2052, Australia\\
$^{2}$Sydney Institute of Astronomy(SIfA), School of Physics, University of Sydney, NSW 2006, Australia\\
$^{3}$Stellar Astrophysics Centre, Dept. of Physics and Astronomy, Aarhus University, Ny Munkegade, DK- 8000 Aarhus C, Denmark.\\
$^{4}$Institute for Astronomy, University of Hawai`i, 2680 Woodlawn Drive, Honolulu, HI 96822, USA\\
$^{5}$Instituto de Astrof\'isica de Canarias, E-38200 La Laguna, Tenerife, Spain \\
$^{6}$Universidad de La Laguna (ULL), Departamento de Astrof\'isica, E-38206 La Laguna, Tenerife, Spain\\
$^{7}$Key Laboratory of Optical Astronomy, National Astronomical Observatories, Chinese Academy of Sciences, Beijing 100101,\\
People's Republic of China\\
$^{8}$Department of Astronomy, Cornell University, Ithaca, NY 14850, USA\\
$^{9}$DTU Space, National Space Institute, Technical University of Denmark, Elektrovej 328, DK-2800 Kgs. Lyngby, Denmark 
}

\date{Accepted XXX. Received YYY; in original form ZZZ}

\pubyear{2019}
\newcommand{\teff}{$T_{\mathrm{eff}}$}
\newcommand{\muhz}{$\mathrm{\mu}$Hz}
\newcommand{\numax}{$\mathrm{\nu}_{\mathrm{max}}$}
\newcommand{\dnu}{$\mathrm{\Delta\nu}$}
\newcommand{\logg}{$\log g$}

\newcommand{\msun}{M$_\mathrm{\odot}$}

\newcommand{\kepler}{\textit{Kepler}}

\newcommand{\kic}{\textit{KIC}}

\newcommand{\gceph}{$\gamma\ $Cep}
\newcommand{\echelle}{\'echelle}
\newcommand{\Echelle}{\'Echelle}
\newcommand{\epstau}{$\epsilon\ $Tau}
\newcolumntype{P}[1]{>{\RaggedRight\hspace{0pt}}p{#1}} % LEft aligned but fixed width columns; called like p{1.5cm}
\begin{document}
\label{firstpage}
\pagerange{\pageref{firstpage}--\pageref{lastpage}}
\maketitle

\begin{abstract}
The study of planet occurrence as a function of stellar mass is important for a better understanding of planet formation. Estimating stellar mass, especially in the red giant regime, is difficult. In particular, stellar masses of a sample of evolved planet-hosting stars based on spectroscopy and grid-based modelling have been put to question over the past decade with claims they were overestimated. Although efforts have been made in the past to reconcile this dispute using asteroseismology, results were inconclusive. In an attempt to resolve this controversy, we study four more evolved planet-hosting stars in this paper using asteroseismology, and we revisit previous results to make an informed study of the whole ensemble in a self-consistent way. For the four new stars, we measure their masses by locating their characteristic oscillation frequency, \numax, from their radial velocity time series observed by SONG. For two stars, we are also able to measure the large frequency separation, \dnu, helped by extended SONG single-site and dual-site observations and new TESS observations. We establish the robustness of the \numax-only-based results by determining the stellar mass from \dnu,  and from both \dnu\ and \numax. We then compare the seismic masses of the full ensemble of  16 stars with the spectroscopic masses from three different literature sources. We find an offset between the seismic and spectroscopic mass scales that is mass-dependent, suggesting that the previously claimed overestimation of spectroscopic masses only affects stars more massive than about 1.6 \msun.
\end{abstract}


%%%%%%%%%%%%%%%%%%%%%%%%%%%%%%%%%%%%%%%%%%%%%%%%%%

%%%%%%%%%%%%%%%%% BODY OF PAPER %%%%%%%%%%%%%%%%%%
\section{Introduction}
\label{sec:intro}
The study of planet occurrence as a function of host star properties, in particular stellar mass, can improve our understanding of planet formation. For this, we need to study potential planet-hosts with a range of stellar masses. However, finding planets around main-sequence stars that are more massive than about 1.4 \msun\ can be challenging, not only for the transit method due to the larger stellar radius \citep{borucki_fresip_1996}, but  particularly for the radial velocity technique, because of the increased line broadening induced by the faster rotation of these stars \citep{johnson_eccentric_2006}. To overcome this, \citet{johnson_eccentric_2006} set out to target intermediate-mass stars in the subgiant and red giant evolution phases, which are more favourable to planet detection using radial velocity measurements. These stars, which they dubbed `retired A-stars', were inferred to be the descendants of main-sequence A- or hot F-type stars.

To find which giants are the descendants of main-sequence A- and hot F-type stars require estimates of stellar mass. However, stellar mass is notoriously difficult to obtain for red giants and late subgiants. Stellar mass is typically estimated by interpolating observed stellar properties such as absolute magnitude, spectroscopy-based metallicity ([Fe/H]), effective temperature (\teff), and surface gravity (\logg) onto stellar model grids \citep{prieto_fundamental_1999, pont_isochrone_2004}. However, the stellar models of a large range of masses converge in the red giant regime of the HR-diagram such that models with different masses and thus, different evolution speeds are within the observed error box. This led \citet{lloyd_retired_2011} to question the inferred masses of the so-called retired A-star sample, suggesting they could be overestimated by up to 50\% (based on a selection of evolved planet-hosting stars from the Exoplanet Orbit Database\footnote{\url{www.exoplanets.org}}, \citealt{wright_exoplanet_2011}). 

Later, \citet{johnson_retired_2013} applied an apparent magnitude limit on their sample of subgiants and benchmarked them against a Galactic stellar population model to show that there was no overestimation in the spectroscopic masses of these evolved planet-hosting stars. The imposed apparent magnitude limit increased the relative number of massive stars ($M \gtrsim$ 1.5 \msun) observed in their target sample, and hence \citet{johnson_retired_2013} argued that this limit partially counteracts the otherwise lower number of massive stars expected from their faster evolution. However, \citet{lloyd_mass_2013} repeated the calculation in \citet{lloyd_retired_2011}, now using apparent magnitude-limited weights for the isochrone integration. From these recalculations, \citet{lloyd_mass_2013} showed that there are fewer massive stars than found in the literature, irrespective of the limit used in the target selection (volume- or magnitude-limit). Meanwhile, \citet{schlaufman_evidence_2013} determined model-independent masses from space velocity dispersions. They found that the velocity dispersions of their subgiant sample were larger than for their main-sequence A0-F5 stars but consistent with their main sequence F5-G5 sample. Hence, they concluded that their evolved planet-hosting stars are less massive than A0-F5 stars, in agreement with \citet{lloyd_retired_2011}. Although not dealing with ensembles like the studies above, \citet{joshua_kelt11} concluded from a comprehensive full system analysis that KELT-11 is indeed a 'retired A-star' with a mass significantly greater than $\sim$ 1.2 \msun. Due to the conflicting results obtained, the debate continued about the true masses of these evolved planet-hosting stars. 

While classical spectroscopically-based mass determinations can be difficult due to the relatively large uncertainties on the spectroscopic parameters like effective temperature, metallicity and surface gravity, recent breakthroughs in asteroseismology have demonstrated that using asteroseismic measurements can provide more precise stellar masses \citep{huber_fundamental_2012, gaulme_testing_2016, huber_asteroseismology_2017, yu2018asteroseismology}, independent of stellar models \citep{stello_oscillating_2008, kallinger_oscillating_2010, chaplin_asteroseismology_2013, basu_asteroseismic_2018}. Thus, asteroseismology is an obvious approach to resolving the dispute over the masses of these evolved planet-hosting stars.

Despite the precision of asteroseismology, the masses of these stars are still contentious. \citet{johnson_physical_2014} made the first attempt to study the only star (HD 185351) in the \kepler\ field that was among the known intermediate-mass evolved planet-hosting stars previously found by radial-velocity (hence amenable to asteroseismic investigation). Unfortunately, only one month of \kepler\ data was available, and no definite conclusion could be made because no unique solution could reconcile all (spectroscopic, seismic, and interferometric) measurements at hand.  However, a follow-up study \citep{hjorringgaard_testing_2017} with a more comprehensive analysis of the asteroseismic data and associated modelling found a unique solution that reconciled all measurements. They concluded that the disputed spectroscopy-based mass was overestimated by about 15\%. \citet{stello_asteroseismic_2017} also found that the spectroscopic masses of seven of the eight evolved planet-hosting stars they studied with the ground-based Stellar Observations Network Group (SONG) telescope \citep{mads_song_2016} were 15--20\%  higher than their corresponding seismic masses. \citet{timwhite_interferometry_2018} determined the masses of 5 evolved planet-hosts based on interferometry and also found the spectroscopic masses from the literature to be 15\% larger than their values. On the other hand, \citet{campante_weighing_2017} and \citet{north_masses_2017} found no apparent difference between the spectroscopic and seismic masses in their sample of stars (not all planet-hosting) observed by K2. Similarly, \citet{ghezzi_2015} found the difference between the spectroscopic and seismic mass scales insignificant compared to the uncertainty in the stellar masses they obtained. 

In this paper, we further investigate the masses of the evolved planet-hosting stars that were previously called into question. For this purpose, we observed four evolved planet-hosting stars for 1--2 weeks in 2018 using the Tenerife node of the SONG telescope. We used the oscillations to estimate the stellar masses following the approach by \citet{stello_asteroseismic_2017}. In addition, we observed one star, \gceph, for two months in 2014 from the SONG telescope at Tenerife and again in 2017 for about three weeks simultaneously from two SONG nodes (Tenerife and Delingha, China). We use the data from these two independent observations to check the robustness of the initial 1-2 week-based SONG data. For one of the stars in our sample, 24 Sex, we verify our findings of the SONG-based seismic masses against the seismic mass obtained from the \textit{Transiting Exoplanet Survey Satellite} (TESS) \citep{ricker_tess_2015}. Finally, we combine the results from our seismic analysis with those of \citet{stello_asteroseismic_2017} and \citet{north_masses_2017} to define an ensemble of 16 stars; this allows us to make the most comprehensive seismic-based analysis of the retired A-star mass controversy to date. 

\section{Target Selection and Observations}
\label{sec:tarsel}
\begin{figure}
	\includegraphics[width=\columnwidth]{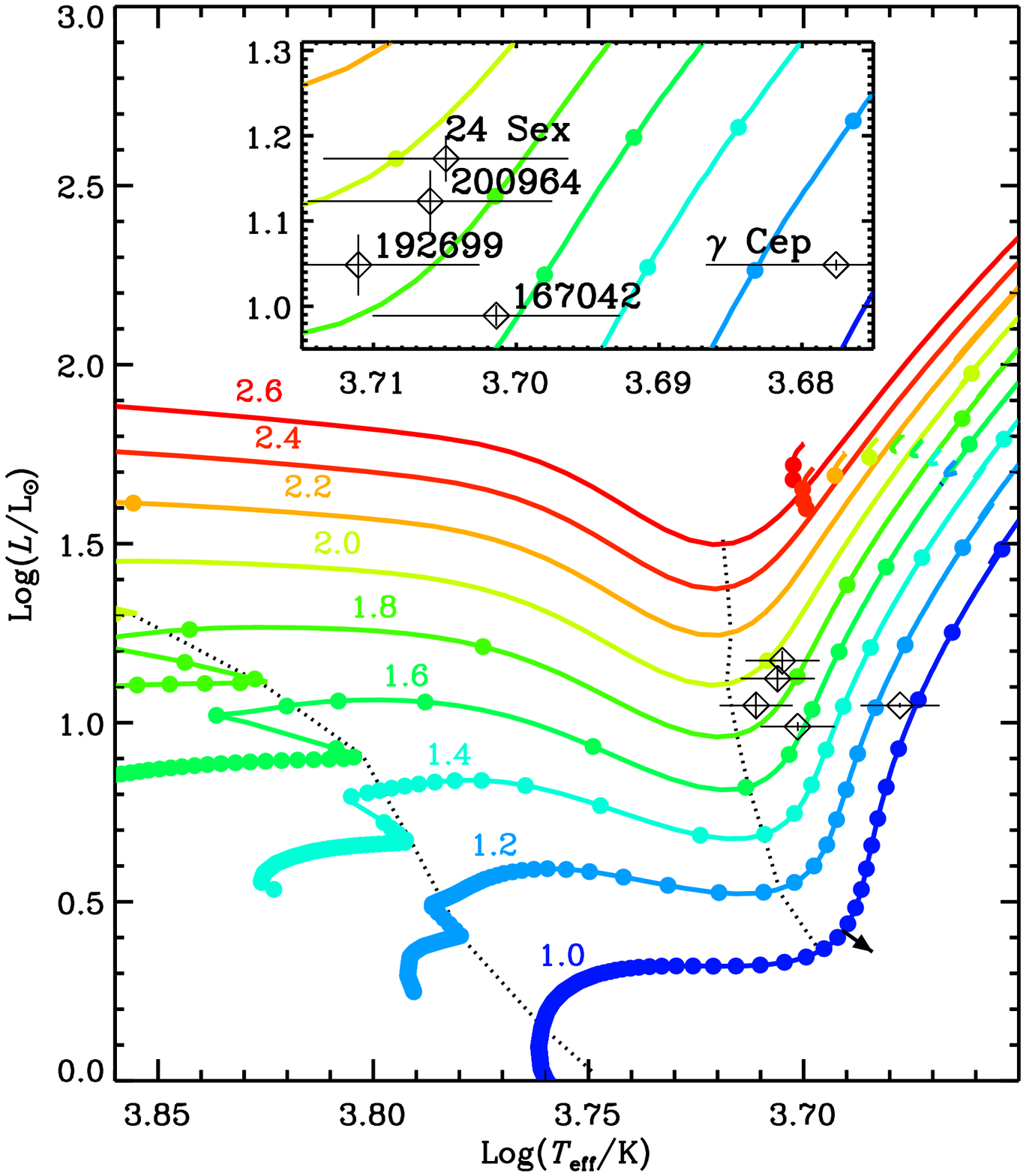}
    \caption{
    The HR-diagram shows the stellar evolution tracks from MESA \citep{paxton_modules_2013} of solar metallicity from \citet{stello_asteroseismic_2013}. The filled dots along each track indicate the likelihood of finding a star in a given state of evolution, each separated by 50 million years in stellar age.  All masses represented in the figure are is solar units. The track shift when the [Fe/H] is increased by 0.2 dex is shown by the black arrow near the bottom of the 1.0 \msun\ red giant branch. The dotted fiducial lines indicate the transitions from the main-sequence to subgiants, and from the rapidly cooling subgiants at roughly the same radius to the rapidly expanding red giants at approximately the same \teff. The planet-hosting targets are represented by diamonds, and the black lines indicate the corresponding uncertainties in their luminosities and effective temperatures. The models within the range of luminosities 1.6 $\lesssim$ log($L/\mathrm{L_{\odot}}$) $\lesssim$ 1.8 are the helium-core burning ones. The inset shows a close-up of the targets on the HR-diagram. 
    }
    \label{fig:hrd}
\end{figure}

We selected our targets from the Exoplanet Orbit Database, which had been the basis for the mass controversy. We used the same selection criteria as \citet{stello_asteroseismic_2017} in effective temperature and luminosity ($L$): $3.65 < \log\ (T_\mathrm{eff}/\mathrm{K}) < 3.75$ (i.e. 4467 K > \teff\ > 5623 K) and $\log\ (L/\mathrm{L_\odot}) > 0.75$ $(L > 5.62\ \mathrm{L_\odot})$. The luminosity of each target was derived from a metallicity-dependent bolometric correction equation \citep[Eq. 18]{alonso_effective_1999} assuming negligible extinction, given the proximity of the targets (see \citealt{stello_asteroseismic_2017} for details). From this initial selection, we chose the four brightest stars with \logg\ > 3 that were not already targeted by \citet{stello_asteroseismic_2017}. Fig.~\ref{fig:hrd} shows our four new targets, along with solar-metallicity stellar evolution tracks from \citet{stello_asteroseismic_2013} derived using MESA \citep{paxton_modules_2013} with dots spaced equally in age.

\noindent\setlength\tabcolsep{2pt}%
\begin{table}
	\centering
	\caption{Observing parameters for targets (all Tenerife except where noted)}
	\label{tab:obslength}
	\begin{threeparttable}
	
	\begin{tabularx}{\linewidth}{P{1.2cm}c*{8}{>{\centering\arraybackslash}X}} % 2 columns, alignment for each
		\toprule
		Star ID &Observation dates & $m_\mathrm{v}$&T$_{\mathrm{exp}}$ & N$_{\mathrm{exp}}$ & R & N$_{\mathrm{night}}^{\mathrm{obs}}$ & N$_{\mathrm{night}}^{\mathrm{span}}$ & $\mathrm{\sigma}_{\mathrm{RV}}$ \\
		 & & & [s] & & & [days] & [days]& [m/s] \\ 
		\midrule
		24 Sex & 05/03--17/03/18 & 6.44 &600 & 404 & 77k & 10 & 12 & 6.70\\
		HD 167042 & 01/06--11/06/18 & 5.95 &900 & 311 & 90k & 10 & 10 & 1.77\\
		HD 192699 & 27/07--11/08/18 & 6.45&1200 & 128 & 90k & 8 & 16 & 3.37\\
		HD 200964 & 17/08--27/08/18 & 6.49 & 1200 & 205 & 90k & 11 & 11 & 2.96\\
		\gceph\ (2014)& 30/08--14/11/14& 3.21 & 180 & 12647 & 90k & 62 & 75 & 2.00\\
	    \gceph\ & 30/10--24/11/17& 3.21 &180 & 860 & 90k & 20 & 23 & 2.60\\
	    \gceph\ (Delingha)& 30/10--22/11/17 & 3.21 & 180 & 2427 & 90k & 21 & 24 & 4.54\\
		\bottomrule
    \end{tabularx}

    $m_\mathrm{v}$ : magnitude \\
	T$_{\mathrm{exp}}$ : exposure time\\
	N$_{\mathrm{exp}}$ : number of exposures\\
	R : spectrograph resolution \\
	N$_{\mathrm{night}}^{\mathrm{obs}}$: number of observation nights \\
	N$_{\mathrm{night}}^{{\mathrm{span}}}$: length of time series \\
	$\mathrm{\sigma}_{\mathrm{RV}}$ : median radial velocity precision \\
	\end{threeparttable}
\end{table}

We used the SONG nodes in Tenerife \citep{andersen_hardware_2014, grundahl_first_2017} and Delingha \citep{deng_song_2012} for the observations. Observations made at Tenerife used the \echelle\ spectrograph of the robotic 1-metre Hertzsprung SONG telescope operated in a fully automated mode \citep{mads_automated_2019}. Observations made at Delingha used a similar spectrograph, but with a slightly shorter spectral range. The operation of the Delingha telescope was not automated, and the observations were carried with an observer present. An iodine cell was used at both nodes for precise wavelength calibration. 

The four new stars in our sample were observed for about 10 days from March to August 2018. In addition, we observed \gceph\ (which was observed by \citealt{stello_asteroseismic_2017} for 13 days) for a period of 75 days from August to November 2014, and using the SONG telescopes from Tenerife and Delingha simultaneously for 24 days from October to November 2017. We combined the dual-site data by shifting each series to a common radial velocity zero-point. The observing parameters are listed in Table~\ref{tab:obslength}. 

\begin{figure*}
	\includegraphics[width=\textwidth]{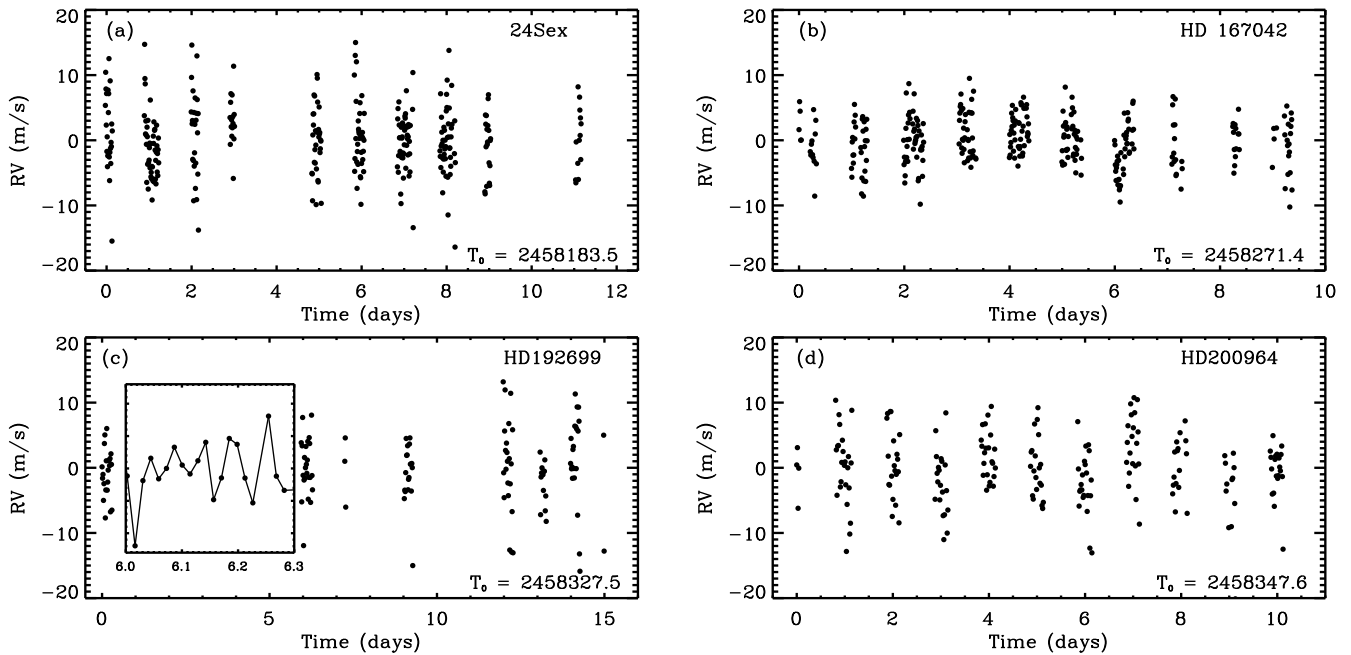}
    \caption{Radial velocity time series for the four new evolved planet-hosting stars studied in this paper. For HD 192699, a single night of observations is shown in the inset. T$_{0}$ is the time (BJD) of the first data point. The data can be acquired from the SONG Data Archive (SODA) or from the author upon request.
    }
    \label{fig:timeseries}
\end{figure*}

\begin{figure}
	\includegraphics[width=\columnwidth,keepaspectratio]{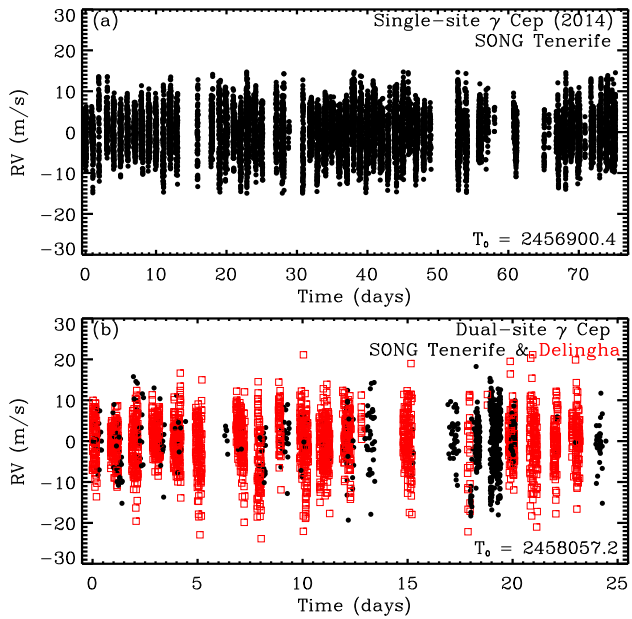}
    \caption{(a) Radial velocity time series for single-site \gceph, which was observed for a period of 75 days from the SONG node at Tenerife. T$_{0}$ is the time (BJD) of the first data point.  (b) Combined radial velocity time series for the dual-site \gceph\ observations. The filled black circles represent the data from Tenerife while unfilled red squares represent the data from Delingha. The data can be acquired from the SONG Data Archive (SODA) or from the author upon request.
    }
    \label{fig:gceph_timeseries}
\end{figure}

The extraction of 1-D spectra and the calculation of radial velocities used the same method as \citet{grundahl_first_2017}. The 1-D spectra were extracted with a pipeline written in Python using C++ routines from \citet{ritter_fast_2014} based on the optimal extraction method by \citet{piskunov_new_2002}. The radial velocities were then calculated following the approach by \citet{butler_attaining_1996} implemented in the \textit{iSONG} software \citep{antoci_searching_2013, grundahl_first_2017}. The radial velocity time series were passed through a high-pass filter with a cutoff frequency of $\sim$ 3 \muhz\ to prevent power leakage in the frequency range of stellar oscillations due to the presence of any slow-moving trends in the data. The final time series after performing a 3$\sigma$-clipping are shown in Figs.~\ref{fig:timeseries} and \ref{fig:gceph_timeseries}.

The radial velocity variations are typically about $\pm$ 10 m/s and dominated by the oscillations as seen in the inset showing a single-night close-up for HD 192699 (Fig.~\ref{fig:timeseries}c).
The radial velocity time series for the single-site (2014) data and the dual-site data for \gceph\ are shown in Fig.~\ref{fig:gceph_timeseries}.

\begin{figure}
	\includegraphics[width=\columnwidth]{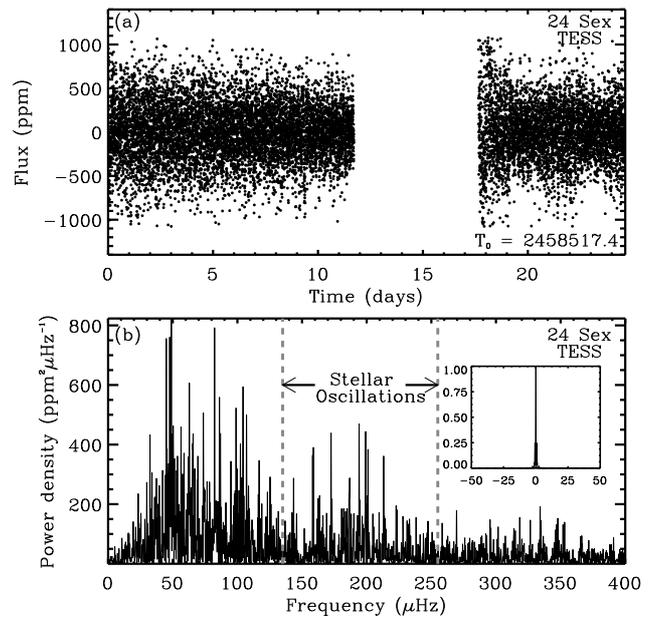}
    \caption{(a) TESS Light curve of 24 Sex. A high-pass filter of $\sim$ 50 \muhz\ is applied. T$_{0}$ is the time (BJD) of the first data point. The data used here can be obtained from \url{http://dx.doi.org/10.17909/t9-fnwn-cr91}. (b) Corresponding power density spectrum. The spectral window is in the inset.}
    \label{fig:tess}
\end{figure}

We also analysed high-precision photometric data from TESS for one of our four new stars, 24 Sex. This star was observed in 2-min cadence in Sector 8 from 2 to 27 February 2019. We downloaded the data from MAST\footnote{\url{https://mast.stsci.edu/portal/Mashup/Clients/Mast/Portal.html}} and used the corrected light curve (PDCMAP) for our analysis. The photometric time series was treated in a similar way to the radial velocity time series, the only exception being the application of a  high-pass filter with a cutoff frequency of $\sim$ 50 \muhz\ due to the larger granulation noise levels at lower frequencies for photometric observations. The high-pass filtered time series of the TESS data for 24 Sex is illustrated in Fig.~\ref{fig:tess}a. 

\section{Measuring \numax\ and its uncertainty}\label{sec:numax}

\begin{figure*}
	\includegraphics[width=\textwidth]{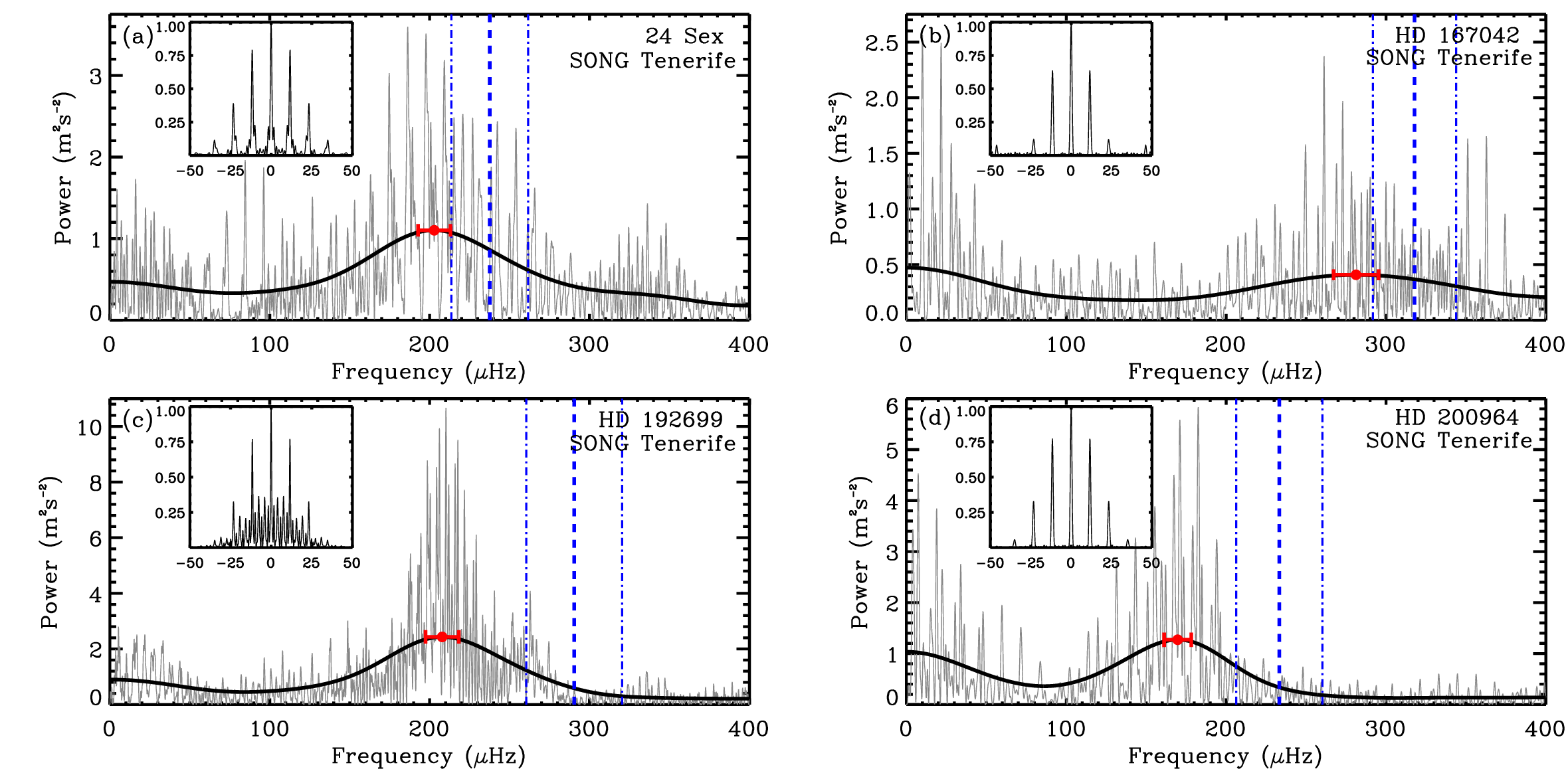}
    \caption{Power spectra of the four new planet-hosting stars observed by SONG. The thick black line is the power spectrum smoothed using a Gaussian of width 4-\dnu. The red dot and the 1$\sigma$ error bars show the observed \numax. The thick dashed blue line represents the \numax\ predicted from Eq.~\ref{eq:scaling} using the spectroscopic \teff\ and mass from the Exoplanet Orbit Database (Table~\ref{tab:data}, column 8) and the thinner dot-dash blue lines represent the corresponding uncertainty. The spectral windows are plotted in the insets. 
    }
    \label{fig:ft}
\end{figure*}

Following \citet{stello_asteroseismic_2017}, we used the same method as \citet{huber_automated_2009} to locate the frequency of maximum oscillation power, \numax. In detail, we calculated the power spectra of the radial velocity time series using a discrete weighted Fourier transform. The resulting power spectra are shown in Figs.~\ref{fig:tess}b, \ref{fig:ft} and \ref{fig:combined_pds}. 
Using a large frequency separation, \dnu, estimated from the approximate \dnu\ $-$ \numax\ relation \citep[Eq. 1]{stello_relation_2009}, we smoothed the power spectrum with a 4\dnu\ wide Gaussian. The highest point of the heavily smoothed power spectrum was taken as \numax\ (Fig.~\ref{fig:ft}, red dot), and the values are tabulated in Table~\ref{tab:data} (column 9). 

We tested that the exact choice of the Gaussian smoothing width did not significantly affect our final \numax\ determination. The test was conducted by varying the Gaussian width by $\pm$ 50\% (corresponding to 2\dnu), which changed the final \numax\ estimate by no more than $\pm$ 2\% for three out of the four new stars in our sample. For one star HD 167042, the change was $\pm$ 5\% due to its broader excess power in the oscillation spectrum. We also note that correcting for any power-loss due to the averaging effect on oscillations during the integration time, like in the case of \kepler\ long-cadence data \citep[Eq. 1]{murphy_examination_2012}, only changes the \numax\ by $\lesssim$ 1\%. Further, \citet{stello_asteroseismic_2017} note that their inferred \numax\ values did not change significantly (less than 1\%) whether or not one takes the stellar background noise into account (see \citet{stello_asteroseismic_2017} for details). This is because the background is very low in radial velocity measurements.

\noindent\setlength\tabcolsep{4pt}
\begin{table*}
	\centering
	\caption{Observed parameters of the evolved planet-hosting stars}
	\label{tab:data}
	\begin{threeparttable}
	\begin{tabularx}{\linewidth}{lc*{9}{>{\centering\arraybackslash}X}}
		\toprule
		& \multicolumn{5}{c}{Literature} &\multicolumn{2}{c}{Derived} & \multicolumn{2}{c}{Asteroseismology}\\
		\cmidrule(l{2pt}r{2pt}){2-6} \cmidrule(l{2pt}r{2pt}){7-8} \cmidrule(l{2pt}r{2pt}){9-10}\\
		Star name & \logg\  & \teff\ & [Fe/H] & $\pi$ & $M$ & $L$ & $\mathrm{\nu}_{\mathrm{max,pre}}$ & $\mathrm{\nu}_{\mathrm{max,obs}}$ & $M$\\
		 & [dex] & [K] & [dex] & [mas] & [\msun] & [$\mathrm{L_{\sun}}$] & [$\mathrm{\mu Hz}$] & [$\mathrm{\mu Hz}$] &[\msun]\\
		(1) & (2)\tnote{a} & (3)\tnote{a} & (4)\tnote{a} & (5)\tnote{b} & (6)\tnote{a} & (7)\tnote{c}& (8)  & (9) & (10)\tnote{d} \\
		\midrule
		24 Sex & 3.40 $\pm$ 0.13 & 5069 $\pm$ 62 & -0.01 $\pm$ 0.05 & 12.91 $\pm$ 0.38 & 1.81 $\pm$ 0.08 & 14.90 $\pm$ 0.92 & 238 $\pm$ 24 & 203 $\pm$ 10 & 1.55 $\pm$ 0.16 \\
		HD 167042 & 3.35 $\pm$0.18 & 5028 $\pm$ 53 & 0.03 $\pm$ 0.04 & 19.91 $\pm$ 0.26 & 1.63 $\pm$ 0.06  & 9.75 $\pm$ 0.27 & 318 $\pm$ 26 & 281 $\pm$ 14 & 1.44 $\pm$ 0.13 \\
		HD 192699 & 3.45 $\pm$ 0.07 & 5141 $\pm$ 20 & -0.2 $\pm$ 0.02 & 15.24 $\pm$ 0.57 & 1.58 $\pm$ 0.04 & 11.18 $\pm$ 0.92 & 290 $\pm$ 30 & 208 $\pm$ 10  & 1.13 $\pm$ 0.13 \\
		HD 200964 & 3.41 $\pm$ 0.08 & 5082 $\pm$ 38 & -0.2 $\pm$ 0.03 & 13.85 $\pm$ 0.52 & 1.57 $\pm$ 0.06 & 13.28 $\pm$ 1.09 & 233 $\pm$ 27 & 170 $\pm$ 8 & 1.14 $\pm$ 0.14\\
		\bottomrule
	\end{tabularx}
	\begin{tablenotes}
	     \item[a]{Source: Exoplanet Orbit Database, which refers to \citet{mortier_new_2013}. Similar to \citet{stello_asteroseismic_2017}, we assume $\sigma_\mathrm{T_\mathrm{eff}}$ = 100 K and $\sigma_\mathrm{[Fe/H]}$ = 0.1 dex to derive columns 7-8 and 10 instead of the quoted uncertainties in \teff\ and [Fe/H] \citep{thygesen_dteff_2012}.}  
	     \item[b]{Source: \textit{Hipparcos} \citep{leeuwen_hipparcos_2007}}
	     \item[c]{To be conservative, we used the largest of the two asymmetric errors obtained from \textit{isoclassify}.}
	     \item[d]{\numax-only based asteroseismic masses}
	\end{tablenotes}
	\end{threeparttable}
\end{table*}

\subsection{Estimating \numax\ uncertainty} \label{numax_err}
\citet{stello_asteroseismic_2017} adopted a 15\% assumed \numax\ uncertainty based on their investigation of the observations of $\xi\ $Hya obtained using the Coralie spectrograph on the Euler Telescope at La Silla (which has a similar performance as SONG; \citealt{frandsen_detection_2002}). We are now in position to check this assumption using the longer SONG time series for two of the planet-hosting stars reported by \citet{stello_asteroseismic_2017}: the 75-day long \gceph\ data presented in Fig.~\ref{fig:gceph_timeseries}a, as well as the 110-day long \epstau\ data from \citet{arentoft_asteroseismology_2019}. This allows us to divide these long series into shorter segments, each similar in length to those of our main sample of stars (about 10 days). By measuring the scatter in \numax\ across segments, we can get a realistic estimate of the uncertainty in \numax. This approach is essentially the same as used by \citet{stello_asteroseismic_2017} (with the $\xi\ $Hya data). However, in our case, the instrumentation and the data reduction approach are identical to that of our shorter observation data sets. 

We split the 75-day long single-site \gceph\ time series into segments of 10 days and measure their \numax, treating them as described in Sec.~\ref{sec:numax}. We observe a \numax\ scatter of 2.5\% across these segments. For the 110-day \epstau\ data, we found a \numax\ scatter of 5\% also using 10-day segments. Based on the above test on \gceph\ and \epstau, we adopt a 5\% \numax\ uncertainty for our four new targets, which is also a typical uncertainty for \numax\ from photometry (e.g., \citealt{huber_2011}). 

We note that our adopted 5\% \numax\ uncertainty is three times smaller than the 15\% \numax\ uncertainty estimated by \citet{stello_asteroseismic_2017} from their analysis of the $\xi\ $Hya radial velocity time series. $\xi\ $Hya is in a different phase of evolution (secondary clump star) and oscillates at much lower frequencies compared to \gceph\ or \epstau. As a result, it has a relatively wide envelope of oscillation power \citep{yu2018asteroseismology}, and also the data is not densely sampled, leading to a much lower signal-to-noise ratio. These factors may contribute to the larger intrinsic \numax\ scatter. In Sec.~\ref{sec:mdiff}, we adopt the mass estimates by \citet{stello_asteroseismic_2017} for our ensemble analysis, using our newly derived 5\% uncertainties. We, therefore, provide an updated summary of the results from \citet{stello_asteroseismic_2017} with this fractional uncertainty in Table~\ref{tab:stello}.

\begin{table}
    \centering
    \caption{Updated Results from \citet{stello_asteroseismic_2017}}
    \label{tab:stello}
    \begin{threeparttable}    
    \begin{tabular}{ccc}
    \toprule
     Star name & \numax\ & $M$ \\
     & (\muhz) & (\msun)\\
     (1) & (2) & (3) \\ 
     \midrule
     \epstau\ & 56.9 $\pm$ 2.9 & 2.40 $\pm$ 0.22 \\
     $\mathrm{a\ }$Gem & 84.5 $\pm$ 4.2 & 1.73 $\pm$ 0.17 \\ 
     18 Del & 112 $\pm$ 6 & 1.92 $\pm$ 0.19\\
     \gceph\ & 185 $\pm$ 9 & 1.32 $\pm$ 0.12\\
     HD 5608 & 181 $\pm$ 9 & 1.32 $\pm$ 0.13\\
     $\mathrm{\kappa\ }$CrB & 213 $\pm$ 11  & 1.40 $\pm$ 0.12\\
     6 Lyn &  183 $\pm$ 9 & 1.37 $\pm$ 0.14 \\
     HD 210702 &  223 $\pm$ 11 & 1.47 $\pm$ 0.14\\
     \bottomrule
\end{tabular}
    \end{threeparttable}
\end{table}

\section{Deriving Stellar Masses}
To calculate stellar seismic mass from the observed \numax, we used the following scaling relation \citep{brown_detection_1991, kjeldsen_amplitudes_1995}:
\begin{equation}
    \frac{\nu_\mathrm{max}}{\mathrm{\nu}_\mathrm{max,\sun}} \simeq \frac{M}{\mathrm{M}_\mathrm{\sun}} \left(\frac{T_\mathrm{eff}}{\mathrm{T}_\mathrm{eff,\sun}}\right)^{3.5}\left(\frac{L}{\mathrm{L}_\mathrm{\sun}}\right)^\mathrm{-1}.
	\label{eq:scaling}
\end{equation}
Here we used $\mathrm{\nu}_\mathrm{max,\sun}$ = 3090 \muhz\ and T$_\mathrm{eff,\sun}$ = 5777K \citep{huber_automated_2009} to be consistent with \citet{stello_asteroseismic_2017}. We used \textit{isoclassify}\footnote{\url{https://github.com/danxhuber/isoclassify}} \citep{huber_asteroseismology_2017} to compute the luminosity of the stars in our sample using the spectroscopic \teff\ from the Exoplanet Orbit Database (Table~\ref{tab:data}, Column 3), \textit{Hipparcos}\footnote{For brighter stars (G < 5), Gaia DR2 parallaxes are known to have larger uncertainities and significant systematic errors due to calibration issues \citep{drimmel_gaia_errors_2019}. Four of the stars in our ensemble study in Sec.~\ref{sec:mdiff} have G < 5. In addition, $\mathrm{\beta}\ $Gem does not have a Gaia DR2 parallax measurement. For the rest of the stars, we find the \textit{Hipparcos} parallaxes to be in good agreement with the Gaia parallaxes. Therefore, we use \textit{Hipparcos} parallaxes instead of Gaia, for consistency.} parallax (Table~\ref{tab:data}, Column 5) and Tycho \textit{V}$_\mathrm{T}$ photometry as inputs. We set the \textit{dustmap} parameter to `allsky', which enables the use of a combination of reddening maps from \citet{drimmel_three-dimensional_2003}, \citet{marshall_modelling_2006}, \cite{green_three-dimensional_2015} and \citet{ bovy_galactic_2016} implemented in the \textit{mwdust} package by \citet{bovy_galactic_2016}. The luminosity\footnote{A brief discussion on the reliability of the \textit{isoclassify}-based luminosities is provided in Sec.~\ref{lum_compare}.} and the seismic mass are tabulated in Table~\ref{tab:data} (columns 7 and 10). 

We note that the location of the seismic signal predicted from the same scaling relation (Eq.~\ref{eq:scaling}) using the spectroscopic \teff\ and masses from the Exoplanet Orbit Database is consistently larger than the observed \numax\ (Fig.~\ref{fig:ft}, dashed blue line). The predicted \numax\ is tabulated in Table~\ref{tab:data} (column 8). Likewise, the seismic masses based on \numax\ (through Eq.~\ref{eq:scaling}) are lower than their spectroscopic counterparts for all the four new stars in our sample. 

\section{Large frequency separations of \gceph\ and 24 Sex}\label{sec:gceph}

Support for our \numax-based masses could come from measurements of masses from the frequency separation between overtone modes, \dnu, which scales with the square root of the mean stellar density. Hence,
\begin{equation}
    \frac{\Delta\nu}{\mathrm{\mathrm{\Delta\nu}_\mathrm{\odot}}} \simeq \left(\frac{M}{\mathrm{\mathrm{M}_\mathrm{\odot}}}\right)^\mathrm{0.5}\left(\frac{L}{\mathrm{L_\mathrm{\odot}}}\right)^\mathrm{-0.75}\left(\frac{T_\mathrm{eff}}{\mathrm{T}_\mathrm{eff, \odot}}\right)^3.
    \label{eqn:dnu}
\end{equation}
This provides two different measurements of stellar masses from \numax\ and \dnu\ to check if our results are consistent. We can also combine \dnu\ (Eq.~\ref{eqn:dnu}) with \numax\ (Eq.~\ref{eq:scaling}) to give a mass with very little \teff\ dependence,
\begin{equation}
    \frac{M}{\mathrm{\mathrm{M}_\mathrm{\odot}}} \simeq \left(\frac{\nu_\mathrm{max}}{\mathrm{\nu_\mathrm{max,\odot}}}\right)^3\left(\frac{\Delta\nu}{\mathrm{\Delta\nu_\mathrm{\odot}}}\right)^{-4}\left(\frac{T_\mathrm{eff}}{\mathrm{T}_\mathrm{eff, \odot}}\right)^\mathrm{1.5},
    \label{eqn:dnumax}
\end{equation}
making the results less sensitive to systematic uncertainties in \teff.

For both Eqs.~\ref{eqn:dnu} and ~\ref{eqn:dnumax}, it is known that one needs to apply a correction to \dnu\ in order to obtain a correct mass \citep{sharma_2016_dnucorr}. This comes from the fact that Eq.~\ref{eqn:dnu} is an approximate relation, and stellar models can give a more exact relation for a given star. Here we use the correction software \textit{asfgrid}\footnote{\url{https://ascl.net/1603.009}} by \citet{sharma_2016_dnucorr} to make the appropriate corrections. For the targets selected using our selection criteria (Sec.~\ref{sec:tarsel}), the correction is usually below 2\%. 

Because the SONG observations of our four new stars are short and single-site, \dnu\ cannot be determined. However, we have long enough time series for \gceph\ (both single-site and dual-site) from SONG and for 24 Sex from TESS to measure their \dnu.

\subsection{\gceph} \label{sec:gceph_dnu}
\begin{figure}
	\includegraphics[width=\columnwidth]{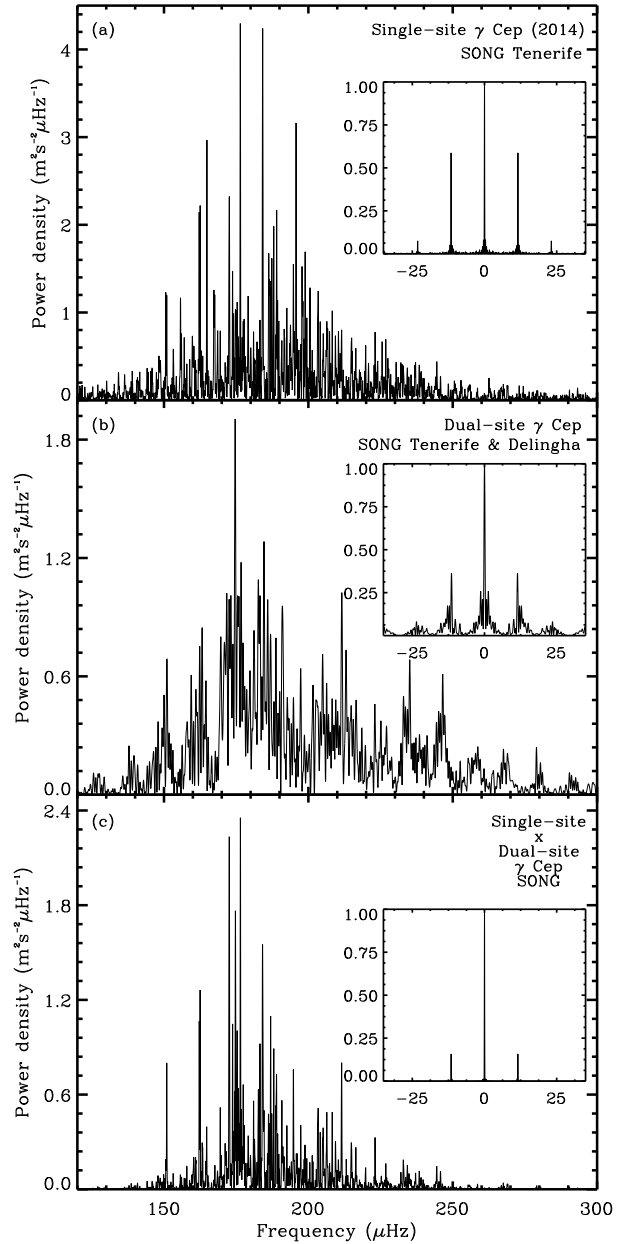}
    \caption{(a) Power density spectrum of single-site \gceph\ observations (b) Power density spectrum of dual-site \gceph\ data (c) Combined power density spectrum of both \gceph\ spectra 
    }
    \label{fig:combined_pds}
\end{figure}

Despite the difficulty of measuring \dnu\ in red giants from ground-based data, our single-site and dual-site data of \gceph\ provide an opportunity to do so. For this purpose, we combined those two data sets by multiplying their respective power density spectra, thus retaining the peaks similar in both spectra while reducing the power of those that are not in common. The resulting power density spectra are shown in Fig.~\ref{fig:combined_pds}.  

\begin{figure}
	\includegraphics[width=\columnwidth]{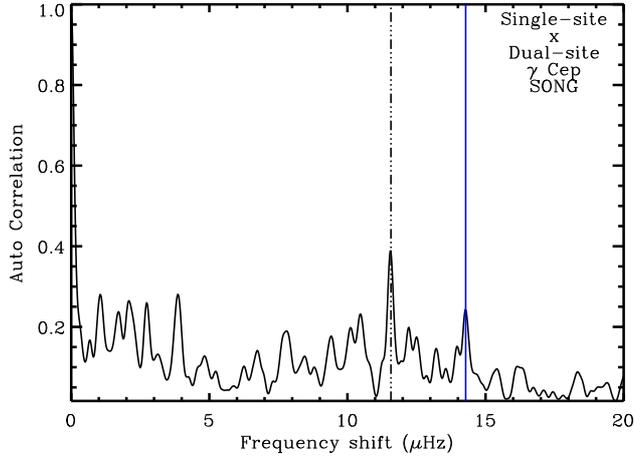}
    \caption{Autocorrelation of the combined power density spectra for \gceph. The dash-dot line represents the daily alias of 11.574 \muhz\ (1 cycle per day). The solid blue line represents the observed \dnu.
    }
    \label{fig:autocorr}
\end{figure}

We performed an autocorrelation on the combined power density spectrum to search for regularity. The peak at the frequency shift for which the correlation is the strongest in the vicinity of the \dnu\ predicted from the \dnu\ - \numax\ relation \citep{stello_relation_2009}, is taken as the \dnu\ peak, and its FWHM gives a conservative uncertainty in \dnu. For \gceph, we obtained a \dnu\ of 14.28 $\pm$ 0.58 \muhz, as can be seen from Fig.~\ref{fig:autocorr}.

Although the autocorrelation allows us to detect \dnu, it does so only marginally and does not give any information about where the underlying modes are located in the spectrum. To investigate the regularity in the power density spectrum further, we divided it into segments of length equal to a trial \dnu\ and stacked them on top of one another. When the trial \dnu\ corresponded to the correct large frequency separation of the stellar oscillations, modes of the same degree aligned vertically with each other. This diagram, known as an \echelle\ diagram \citep{grec1983, bedding_echelle_2010}, allowed us to clearly see which \dnu\ provided alignment (a repeated pattern) and showed the absolute location of the aligned peaks. We use the \textit{echelle}\footnote{\url{https://pypi.org/project/echelle/}} module \citep{danhey_echelle} to plot the \echelle\ diagrams and test the trial \dnu\ for which the peaks align vertically. From \kepler\ data, we know there is a correlation between \dnu\ and the location of the aligned peaks in the \echelle\ diagram \citep{white_calculating_aseismo_diagrams_2011}, which is tighter for red giants (see also \citealt{bedding_echelle_2010}, \citealt{huber_2010_kep}, and \citealt{mosser_2010_corot}) compared to less evolved stars. Hence, the \dnu\ that we find needs to agree with the correct location of the aligned peaks. 

\begin{figure}
	\includegraphics[width=\columnwidth]{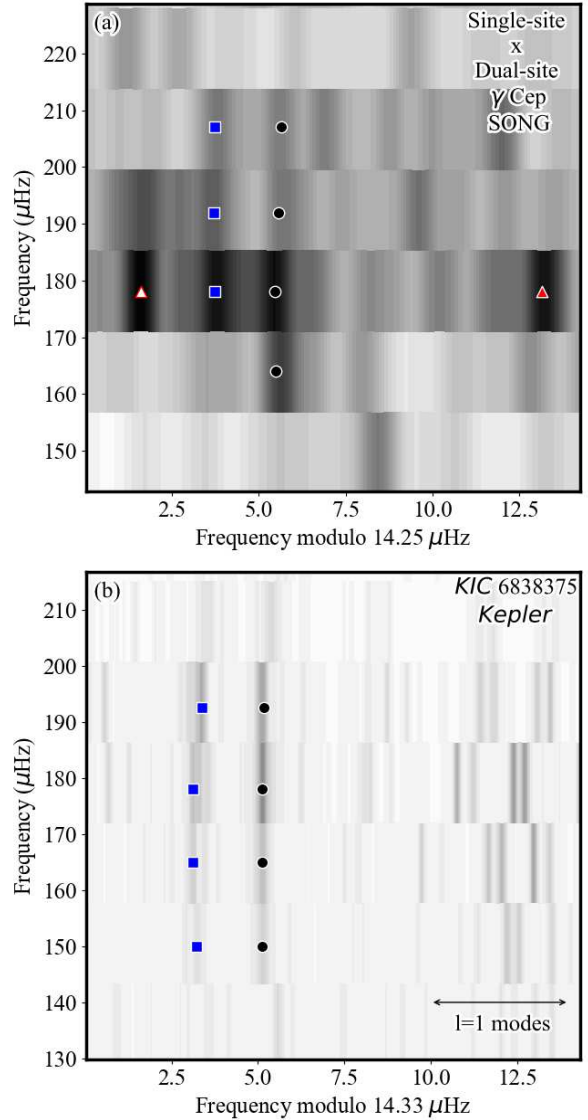}
    \caption{(a) \Echelle\ diagram of \gceph\ computed from the combined smoothed power density spectrum. The filled black circles mark the radial ($l=0$) mode frequencies. The filled blue squares represent the quadrupole ($l=2$) modes. The filled red triangle represents a dipole ($l=1$) mode while the white-filled red triangle represents its alias. Only the modes, which could be clearly distinguished based on their strength and location, are marked. The approximate frequencies corresponding to these modes are provided for reference in Table~\ref{tab:approxfreq} (columns 1 and 2). (b) \Echelle\ diagram of the \kepler\ star \kic\ 6838375, which has a \dnu\ similar to \gceph. Here we mark the region where the strongest dipole modes fall.
    }
    \label{fig:echelle}
\end{figure}

For \gceph, we tested values of \dnu\ from 0 to 20 \muhz. We found that the peaks stacked neatly on top of one another when \dnu\ = 14.25 \muhz\ (Fig.~\ref{fig:echelle}a), which is consistent with our results from the autocorrelation. For comparison, we plot the \echelle\ diagram of the \kepler\ star \kic\ 6838375, which has a similar \dnu\ and \numax\ as \gceph\ \citep{yu2018asteroseismology} (Fig.~\ref{fig:echelle}b). The long continuous time-base of the \kepler\ data enables us to see the oscillations and identify the modes clearly. We find that the \dnu\ observed for \gceph\ creates an \echelle\ similar to that of the representative star observed by \kepler\ (e.g. aligned peaks at similar locations), except at much lower resolution (due to the shorter time series) and with alias peaks present (due to the non-continuous data of SONG). The latter makes it difficult to determine with certainty which of the peaks in the dipole region are real or aliases. We find one peak that is probably real based on its strength and the location in the \echelle\ (red triangle) as well as the location of the peak that we identify as its alias (white triangle). The approximate frequencies for the individual mode frequencies extracted from the \echelle\ diagram are listed in Table~\ref{tab:approxfreq}.

\noindent\setlength\tabcolsep{4pt}%
\begin{table}
    \centering
    \caption{Approximate frequencies of individual modes extracted from the \echelle\ diagrams of \gceph\ and 24 Sex}
    \label{tab:approxfreq}
    \begin{threeparttable}    
    \begin{tabularx}{\linewidth}{c*{4}{>{\centering\arraybackslash}X}} % 2 columns, alignment for each
    \toprule
    \multicolumn{2}{c}{\gceph} & \multicolumn{2}{c}{24 Sex}\\
    \cmidrule(l{2pt}r{2pt}){1-2} \cmidrule(l{2pt}r{2pt}){3-4}\\
     Frequency & Degree & Frequency & Degree \\
     (\muhz) & &(\muhz) & \\
     (1) & (2) & (3) & (4)\\ 
     \midrule
    162.5 & $l=0$ & 158.8 & $l=0$ \\
    176.5 & $l=0$ & 172.5 & $l=0$\\
    190.8 & $l=0$ & 186.8 & $l=0$ \\
    205.1 & $l=0$ & 201.0 & $l=0$\\
    174.7 & $l=2$ & 184.3 & $l=2$ \\
    189.0 & $l=2$ & 199.2 & $l=2$ \\
    203.4 & $l=2$ & 213.1 & $l=2$\\
    184.1 & $l=1$ & 194.1 & $l=1$ \\
     \bottomrule
    \end{tabularx}
    \end{threeparttable}
\end{table}

For \gceph, we obtain a mass of 1.37 $\pm$ 0.15 \msun\ when using \dnu\ alone (from Eq.~\ref{eqn:dnu}) and a mass of 1.20 $\pm$ 0.22 \msun\ when both \dnu\ and \numax\ are used (Eq.~\ref{eqn:dnumax}). These were both in agreement with its previously published \numax-based mass \citep{stello_asteroseismic_2017} even when adopting our new, much smaller \numax\ uncertainties for the \citet{stello_asteroseismic_2017} results (Table~\ref{tab:stello}).

\subsection{24 Sex}
Compared to ground-based observations, it is relatively easy to measure \dnu\ in space-based observations due to the availability of continuous data and hence, lower aliases. The 25-day long TESS data for one of the stars in our sample, 24 Sex, therefore enables us to measure its \dnu.

\begin{figure}
	\includegraphics[width=\columnwidth]{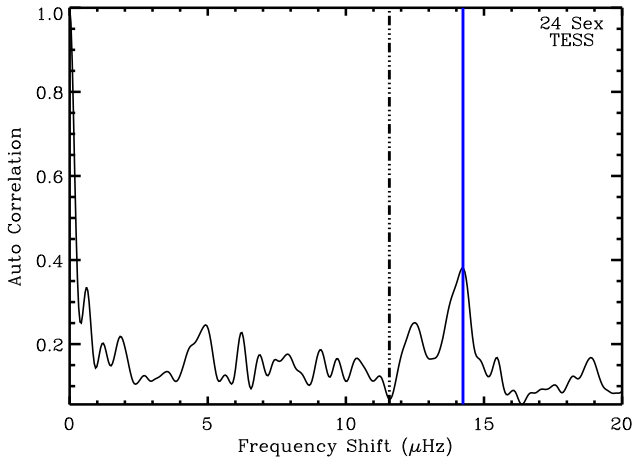}
    \caption{Autocorrelation of the power density spectrum for 24 Sex from TESS data. The dash-dot line represents the daily alias of 11.574 \muhz\ (1 cycle/day), the solid blue line represents the observed \dnu. 
    }
    \label{fig:tess_acf}
\end{figure}

\begin{figure}
	\includegraphics[width=\columnwidth]{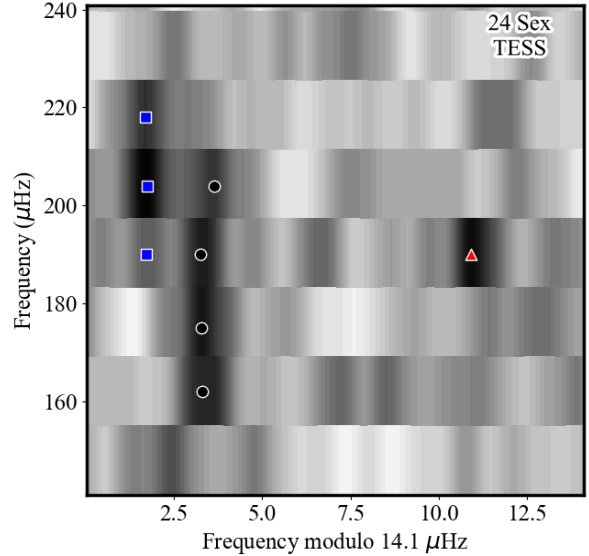}
    \caption{\Echelle\ diagram from the TESS data for 24 Sex. The filled black circles mark the radial mode frequencies ($l=0$), the filled blue squares represent the quadrupole ($l=2$) modes and the filled red triangle represents a dipole ($l=1$) mode. Like for \gceph, only the modes that could be clearly distinguished based on their strength and location are marked. The approximate frequencies corresponding to these modes are provided in Table~\ref{tab:approxfreq} (columns 3 and 4).}
    \label{fig:tess_echelle}
\end{figure}

As for the SONG data of \gceph, we first calculated the autocorrelation of the power density spectrum of the TESS data for 24 Sex. Fig.~\ref{fig:tess_acf} indicates a strong correlation for a frequency spacing of 14.15 $\pm$ 1.23 \muhz.
We find the best vertical alignment of the modes in the \echelle\ diagram for a \dnu\ = 14.10 \muhz\ (Fig.~\ref{fig:tess_echelle}).

We obtain a mass of 1.39 $\pm$ 0.23 \msun\ for 24 Sex based on \dnu\ (Eq.~\ref{eqn:dnu}), and 1.64 $\pm$ 0.38 \msun\ using both \dnu\ and \numax. These results are consistent with the \numax-based mass from SONG that we report in Table~\ref{tab:data} and hence also lower than the spectroscopic mass.

Overall, we see that the \numax-, \dnu- and the `\dnu\ + \numax'-based masses are in good agreement with each other.

\section{Offset between the spectroscopic and seismic masses} \label{sec:mdiff}

From Table~\ref{tab:data}, we see all four new stars presented here show seismic masses lower than the spectroscopic masses from the Exoplanet Orbit Database. This agrees with the results on seven stars from \citet{stello_asteroseismic_2017} but disagrees with the results from \citet{north_masses_2017} and the one star in the \citet{stello_asteroseismic_2017} sample (\gceph), for which the seismic and spectroscopic masses agree.

\begin{figure*}
	\includegraphics[width=\textwidth]{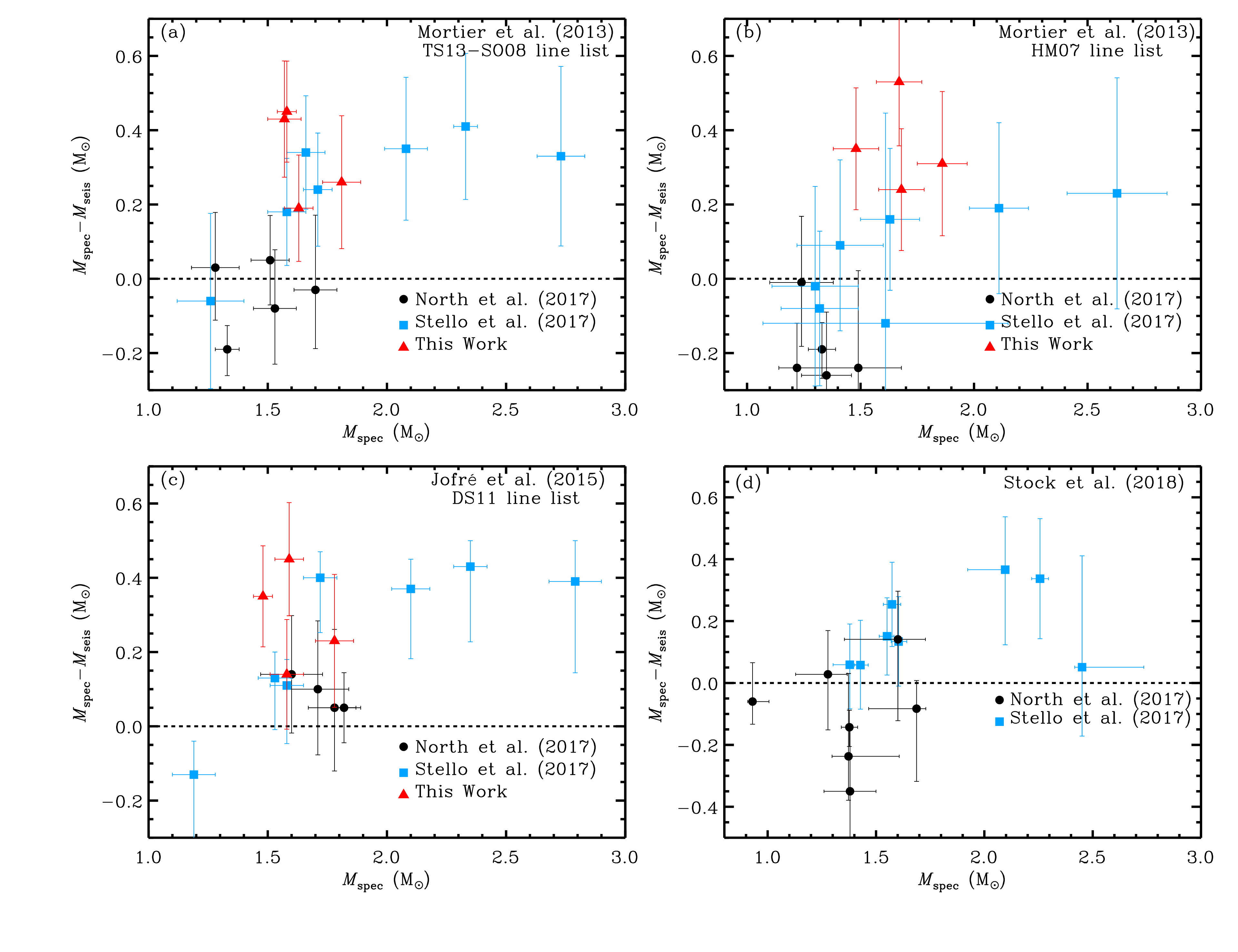}
    \caption{Difference between the spectroscopic and seismic masses plotted as a function of spectroscopic mass from four sources: (a) \citet{mortier_new_2013}, derived using the line lists from \citet[TS13]{tsantaki_linelist_2013} for cooler stars (\teff\ < 5200 K) and \citet[SO08]{sousa_linelist_2008} for hotter stars (16 stars), (b) \citet{mortier_new_2013}, derived using the line list from \citet[HM07]{hekker_linelist_2007} (16 stars), (c) \citet{jofre_stellar_2015}, derived using the iron line list from \citet[DS11]{dasilva_linelist_2011} (15 stars), and (d) \citet{stock_evoltime_2018}, derived using the \teff, [Fe/H] and \logg\ values from \citet{hekker_linelist_2007}  (15 stars). The results from \citet{stello_asteroseismic_2017} with updated error bars  and \citet{north_masses_2017} have also been included. The filled red triangles represent the results obtained from this paper, the filled black circles represent the results from \citet{north_masses_2017}, and the filled blue squares denote the results from \citet{stello_asteroseismic_2017}. 
    }
    \label{fig:massdiff}
\end{figure*}

To further investigate these apparently discrepant results, we combine all the results from the previous papers \citep{stello_asteroseismic_2017, north_masses_2017} with ours, only choosing the stars for which the sources for spectroscopic mass are the same, for consistency\footnote{A list of all the stars in the ensemble and their stellar masses across various literature sources used for this study is provided in Table~\ref{tab:apex_mass}.}. We show in Figs.~\ref{fig:massdiff}a and \ref{fig:massdiff}b the mass difference ($M_\mathrm{seis}$-$M_\mathrm{spec}$) as a function of the spectroscopic mass ($M_\mathrm{spec}$) for the largest sample of stars (16 stars) with a single spectroscopic source that overlap with our combined seismic sample \citep{mortier_new_2013}. 
This combined data shows an interesting trend. The difference between the two mass scales is insignificant for low mass stars, in agreement with the results by \citet{north_masses_2017} (and the lowest mass star by \citealt{stello_asteroseismic_2017}). However, for the more massive stars, the difference between the two scales is pronounced, which agrees with the conclusions made by \citet{stello_asteroseismic_2017}. Here we note that the majority of stars investigated by \citet{north_masses_2017} are of lower mass than those investigated by \citet{stello_asteroseismic_2017}. We observe a sudden increase in the offset between the two mass scales at about 1.6 \msun. Here, we note that \citet{mortier_new_2013} provided two sets of spectroscopic masses derived using different line lists: one set used the \citet{tsantaki_linelist_2013} line list for cooler stars (\teff\ < 5200K) and the \citet{sousa_linelist_2008} line list for the hotter stars in their sample (Fig.~\ref{fig:massdiff}a); the other used the \citet{hekker_linelist_2007} line list, which was specifically made for giants to avoid blends due to atomic and CN lines (Fig.~\ref{fig:massdiff}b). The stellar masses from these two different line lists show a slight deviation in the mass range 1.7--2.1 \msun\ \citep[Fig. 2]{mortier_new_2013}. By comparing Figs.~\ref{fig:massdiff}a and \ref{fig:massdiff}b (same method but different line lists), it is evident that the choice of line list matters, but that the mass-dependent offset relative to the seismic mass occurs in both cases. The increasing offset with mass persists even when we adopt other spectroscopic sources, albeit with fewer stars in common with our seismic sample: \citet{jofre_stellar_2015} (15 stars, Fig.~\ref{fig:massdiff}c) and \citet{stock_evoltime_2018} (15 stars, Fig.~\ref{fig:massdiff}d). \citet{jofre_stellar_2015} derived their spectroscopic masses using the iron line lists from \citet{dasilva_linelist_2011}. \citet{stock_evoltime_2018} did not use line lists directly in their analysis, but used the \teff, [Fe/H] and \logg\ values from \citet{hekker_linelist_2007}. Despite a less clear jump at 1.6 \msun, the comparison with \citet{stock_evoltime_2018} still shows a slight positive trend with increasing $M_\mathrm{spec}$, though barely significant. With a larger sample of 26 stars with seismic data, \citet{stock_evoltime_2018} found a positive offset with a negative slope, but both the offset and slope were compatible with zero. Hence, they concluded the offset to be insignificant.

\citet{johnson_giant_2010} use the stellar masses from the Spectroscopic Properties of Cool Stars (SPOCS) catalog \citep{valenti_spocs_2005} for their calculation of planet occurrence-mass-metallicity correlation. Of the planet-hosting subgiants studied by \citet{johnson_giant_2010} for their planet occurrence-mass-metallicity correlation, the massive stars ($M \gtrsim$ 1.6 \msun) constitute $\sim$ 46\%. Correcting for the observed mass-dependent offset would push the retired A-star sample to smaller masses, which would result in a steeper planet occurrence as a function of stellar mass compared to what was presented by \citet[Eq. 8]{johnson_giant_2010}. We found an overlap of 13 stars between our full seismic sample and the spectroscopic sample of \citet{brewer_spocs_2016}, which is a part of the full SPOCS sample. However, all the stars in the overlap had spectroscopic-based isochrone masses less than 1.7 \msun and unsurprisingly showing no mass-offset correlation. Hence, no conclusion could be drawn. Further investigation with a larger sample in a range of stellar masses from 1 to 3 \msun\ is required to check for the offset between the spectroscopic masses from the SPOCS catalog and the seismic masses. If an offset exists, a recalculation of the planet occurrence-mass-metallicity correlation will be needed. Such investigation is beyond the scope of the current paper and will be performed in future work (Malla et al. in prep.). 

Given that our seismic masses $M_\mathrm{seis}$ plotted in Fig.~\ref{fig:massdiff} are based on \numax, one could suspect that Eq.~\ref{eq:scaling} provides biased results; either because the different quantities that go into the relation (\teff, $L$, \numax) are biased or because the relation itself breaks down. However, \citet{stello_asteroseismic_2017} previously studied the effect of the potential systematics on \numax. They determined the adopted \teff\ was unlikely to be off by enough to affect the \numax\ by such a significant amount as the mass offset we see beyond 1.6 \msun\ (this is also supported by our consistent masses from Eqs.~\ref{eq:scaling}, \ref{eqn:dnu}, and \ref{eqn:dnumax}, given their different dependence on \teff). They noted that a systematic shift in metallicity by 0.1 dex only alters the \numax\ predicted from spectroscopy by 4\% for stars on the red giant branch. They also found it highly unlikely for the \numax\ scaling relations to be off by 15--20\% for red giants. They concluded that the potential systematics only affected the \numax\ by 4-5\%, which is within our adopted uncertainty. Thus, it seems safe to assume the potential systematics in the seismic mass does not cause the observed offset. This has subsequently been supported by the comparison of radii and masses based on Eqs.~\ref{eq:scaling} and \ref{eqn:dnu} with results from Gaia \citep{zinn2019_gaia} and Galactic stellar populations \citep{sharma_k2_hermes_2019}, suggesting even less room for error in Eq.~\ref{eq:scaling}. The sudden jump in stellar rotation speeds, Kraft break, occurs at 1.2 \msun\ \citep{kraft1967}, and thus, the observed offset is unlikely to be associated with this jump in rotational velocities either. 

The transition mass of 1.6 \msun\ for the offset is about the same as the one that separates slow- and fast-evolving stars in the lower red giant branch region, which is where most of our `retired A-star' targets lie. From Fig.~\ref{fig:hrd}, it is clear that massive stars ($M \gtrsim$ 1.6 \msun) evolved much faster (thus spend less time) in the target region. Given the size of the spectroscopy-based uncertainties and the merging of tracks of different masses on the red giant branch, the typical spectroscopic error box can easily encompass low mass (slow and hence more likely) and high mass (fast and hence less likely) evolving tracks at the same time. Therefore, if the evolution speeds are not properly accounted for, the inferred stellar masses can be easily overestimated \citep{lloyd_retired_2011}. Here, we note that \citet{stock_evoltime_2018}, which showed the smallest mass-offset among our comparisons, is the only spectroscopic source that explicitly mention they take the stellar evolution speed into account when estimating the stellar masses. 

\section{Conclusions}
We used radial velocity time series from the ground-based SONG telescopes to determine the asteroseismic masses of four evolved planet-hosting stars that have not previously been investigated using asteroseismology. Our observations are too short to enable the measurement of the large frequency separation or individual mode frequencies. With especially long or less interrupted data for \gceph\ (a star previously reported by \citealt{stello_asteroseismic_2017}) and 24 Sex (a star from our sample that has also been observed by TESS), we were able to establish the robustness of the results that were based on the shorter base-line data by independently estimating the stellar mass from \dnu\ alone and from \dnu\ and \numax\ combined. 

We found an offset between the spectroscopic and seismic masses above a transition mass of 1.6 \msun. Our results are consistent with \citet{north_masses_2017}, who found no offset for less massive stars, and with \citet{stello_asteroseismic_2017}, who found an offset for more massive stars. Our results also agree with the more recent result by \citet{campante_tessrg_2019}, who found a TESS-based seismic mass of 1.23 $\pm$ 0.15 \msun\ against a spectroscopic mass of 2.1 $\pm$ 0.1 \msun\ for the evolved planet-host HD 203949. These results suggest that the spectroscopy-based stellar masses of massive stars ($M \gtrsim$ 1.6 \msun) are prone to overestimation, which implies that planet occurrence increases even more steeply with host star mass, compared to previous estimates \citep{johnson_giant_2010, ghezzi_retired_2018}. 

TESS is currently observing many of these evolved planet-hosting stars, which will enable us to measure their \numax, and possibly \dnu\ and individual mode frequencies. By combining these data with the Gaia DR2 parallax measurements, we should be able to get more precise mass estimates for an even larger sample of previously reported evolved planet-hosts that bracket the mass around the transition mass to further confirm our finding and recalculate the planet occurrence-mass-metallicity correlation towards intermediate-mass stars.

\section*{Acknowledgements}
This research is based on observations made with the SONG telescopes operated on the Spanish Observatorio del Teide (Tenerife) and at the Chinese Delingha Observatory (Qinghai) by the Aarhus and Copenhagen Universities, by the Instituto de Astrof\'{i}s\'{i}ca de Canarias and by the National Astronomical Observatories of China. Funding for the Stellar Astrophysics Centre is provided by The Danish National Research Foundation (Grant agreement no.: DNRF106). D.S. acknowledges support from Australian Research Council. D.H. acknowledges support by the National Science Foundation (AST-1717000).
This research had made use of the Exoplanet Database and the Exoplanet Data Explorer at exoplanets.org. This research also uses data collected by the TESS mission, which is funded by NASA Explorer program and obtained from the Mikulski Archive for Space Telescopes (MAST). STScI is operated by the Association of Universities for Research in Astronomy, Inc., under NASA contract NAS5-26555. Support for MAST for non-HST data is provided by the NASA Office of Space Science via grant NNX13AC07G and by other grants and contracts. 
In addition, this research has made use of NASA's Astrophysics Data System Bibliographic Services. 

\section*{Data Availability}
The SONG data used in this article can be acquired from the SONG Data Archive (SODA) or from the author upon request. The TESS data for 24 Sex used in this article can be obtained from \url{http://dx.doi.org/10.17909/t9-fnwn-cr91}. All the tables in this paper are available on CDS in a machine-readable format. 

%%%%%%%%%%%%%%%%%%%%%%%%%%%%%%%%%%%%%%%%%%%%%%%%%%



\bibliographystyle{mnras}
%\interlinepenalty=10000
\bibliography{ref} % if your bibtex file is called example.bib

%\begin{thebibliography}{99}
%\end{thebibliography}

%%%%%%%%%%%%%%%%%%%%%%%%%%%%%%%%%%%%%%%%%%%%%%%%%%

%%%%%%%%%%%%%%%%% APPENDICES %%%%%%%%%%%%%%%%%%%%%

 \appendix
\section{A note on the reliability of \textit{isoclassify}-based luminosities}\label{lum_compare}

\begin{figure}
	\includegraphics[width=\columnwidth]{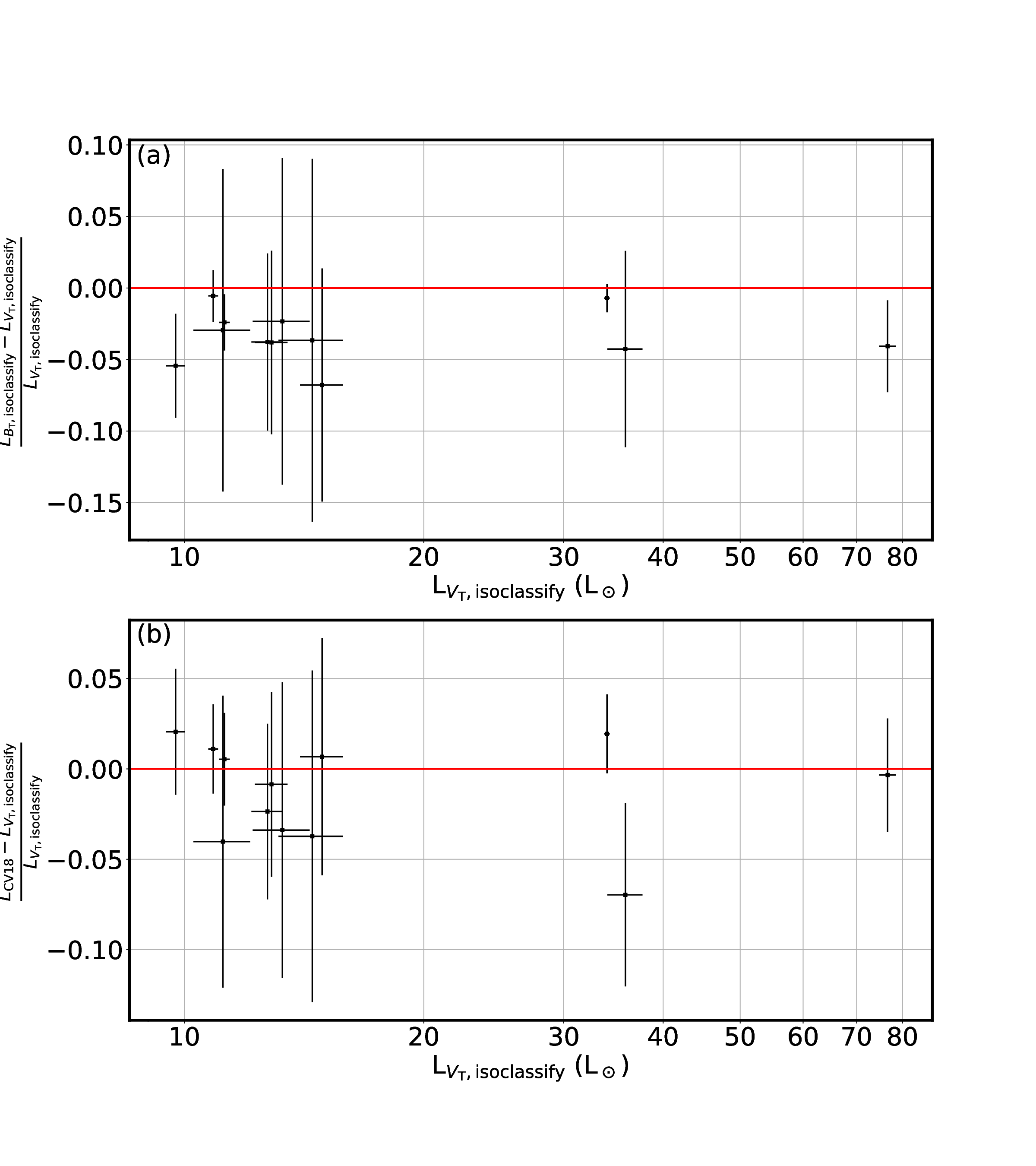}
    \caption{(a) Fractional difference in luminosities derived from \textit{isoclassify} using Tycho $B_\mathrm{T}$ and $V_\mathrm{T}$ photometry as a function of the $V_\mathrm{T}$-based luminosity. The solid red line represents zero difference. (b) Fractional difference in the luminosities derived using \citet{luca_bc_2018} bolometric corrections in the Tycho $V_\mathrm{T}$ passbands using the MARCS models and the luminosity derived from \textit{isoclassify} for the Tycho $V_\mathrm{T}$ as a function of the latter. 
    }
    \label{fig:luca}
\end{figure}

To test the reliability of the luminosities obtained using \textit{isoclassify}, we ran the analysis again using Tycho $B_\mathrm{T}$ photometry instead of Tycho $V_\mathrm{T}$. Fig.~\ref{fig:luca}a shows the fractional difference in the luminosity thus obtained using \textit{isoclassify} based on $B_\mathrm{T}$ ($L_{B_\mathrm{T},\mathrm{isoclassify}}$) and on $V_\mathrm{T}$ photometry ($L_{V_\mathrm{T},\mathrm{isoclassify}}$), as a function of the latter. We note an average difference of ~3.4\% between the two, which is less than our adopted uncertainty of 5\%.

\textit{isoclassify} uses the MIST grids\footnote{\url{http://waps.cfa.harvard.edu/MIST/model_grids.html}} to interpolate bolometric corrections from spectroscopic \teff, surface gravity \logg, metallicity [Fe/H] and extinction $A_V$. \citet{zinn_radius_2019} found the MIST K$_s$-band bolometric fluxes derived using the corresponding bolometric corrections to be consistent with the bolometric corrections from the InfraRed Flux Method, g-band MIST and another K$_s$-band bolometric corrections from \citet{gonzalez_vband_2009} within 4\% (see Fig. 14, \citealt{zinn_radius_2019}). They also showed that MIST bolometric corrections are consistent with each other within 3\% (for i-, g- and r-bands). Therefore, it would seem safe to assume the Tycho $V_\mathrm{T}$ and $B_\mathrm{T}$ bolometric corrections from the MIST grids to be consistent with other sources of bolometric corrections.

As an additional test, we checked how different bolometric corrections for $V_\mathrm{T}$ affect the derived luminosities for our test sample of 12 stars. For these 12 stars, we obtained bolometric corrections from the \citet{luca_bc_2018} tables covering Tycho $V_\mathrm{T}$ passbands. We then used these bolometric corrections to compute the luminosities ($L_\mathrm{CV18}$). Fig.~\ref{fig:luca}b demonstrates the fractional difference in $L_\mathrm{CV18}$ and $L_{V_\mathrm{T},\mathrm{isoclassify}}$ as a function of the latter. We note that the average fractional difference between these two luminosities is ~1.3\%, which is below our adopted uncertainty.

\section{Stellar masses for the ensemble of evolved planet-hosting stars across different sources in literature}
\noindent\setlength\tabcolsep{4pt}%
\renewcommand{\arraystretch}{1.5}
\begin{table*}%[!htbp]
	\centering
	\caption{Stellar masses for the evolved planet-hosting stars used for the ensemble study in Sec.~\ref{sec:mdiff} across various literature sources}
	\label{tab:apex_mass}
	\begin{threeparttable}
	
	\begin{tabularx}{\linewidth}{lc*{4}{>{\centering\arraybackslash}X}}
		\toprule
		\multicolumn{1}{l}{Star name}& \multicolumn{4}{c}{Spectroscopy-based grid-based modelling} & \multicolumn{1}{c}{Asteroseismology}\\
		\cmidrule(l{2pt}r{2pt}){1-1}\cmidrule(l{2pt}r{2pt}){2-5} \cmidrule(l{2pt}r{2pt}){6-6}\\
		 & \citet{mortier_new_2013} & \citet{mortier_new_2013} & \citet{jofre_stellar_2015} & \citet{stock_evoltime_2018} & \\
		 & TS13-SO08 & HM07 & DS11 & & \\
		 & [\msun]& [\msun]& [\msun]& [\msun]& [\msun]\\
		(1) & (2)\tnote{a} & (3)\tnote{b} & (4)\tnote{c} & (5)\tnote{d} & (6) \\
		\midrule
		24 Sex & 1.81 $\pm$ 0.08 & 1.86 $\pm$ 0.11 & 1.78 $\pm$ 0.08 & -- & 1.55 $\pm$ 0.16\tnote{e}\\
		HD 167042 & 1.63 $\pm$ 0.06 & 1.68 $\pm$ 0.1 & 1.58 $\pm$ 0.07 & -- & 1.44 $\pm$ 0.13\tnote{e} \\
		HD 192699 & 1.58 $\pm$ 0.04 & 1.48 $\pm$ 0.1 & 1.48 $\pm$ 0.04 & -- & 1.13 $\pm$ 0.13\tnote{e}\\
		HD 200964 & 1.57 $\pm$ 0.07 & 1.67 $\pm$ 0.1 & 1.59 $\pm$ 0.06 & -- & 1.14 $\pm$ 0.14\tnote{e}\\
		\epstau\ & 2.73 $\pm$ 0.1 & 2.63 $\pm$ 0.22 & 2.79 $\pm$ 0.11 & $\mathrm{2.451}^\mathrm{+0.285}_\mathrm{-0.034}$ & 2.40 $\pm$ 0.22\tnote{f}\\
        $\mathrm{\beta\ }$Gem & 2.08 $\pm$ 0.09 & 1.61 $\pm$ 0.54 & 2.1 $\pm$ 0.08 & $\mathrm{2.096}^\mathrm{+0.018}_\mathrm{-0.173}$ & 1.73 $\pm$ 0.17\tnote{f}\\
        18 Del & 2.33 $\pm$ 0.05 & 2.11 $\pm$ 0.13 & 2.35 $\pm$ 0.07& $\mathrm{2.257}^\mathrm{+0.039}_\mathrm{-0.039}$ & 1.92 $\pm$ 0.19\tnote{f}\\
        \gceph\ & 1.26 $\pm$ 0.14 & 1.3 $\pm$ 0.19 & 1.19 $\pm$ 0.09& $\mathrm{1.379}^\mathrm{+0.054}_\mathrm{-0.077}$ & 1.32 $\pm$ 0.19\tnote{f}\\
        HD 5608 & 1.66 $\pm$ 0.08 & 1.41 $\pm$ 0.19 & 1.72 $\pm$ 0.07 & 1.574 $\pm$ 0.040 & 1.32 $\pm$ 0.13\tnote{f}\\
        $\mathrm{\kappa\ }$CrB & 1.58 $\pm$ 0.08 & 1.32 $\pm$ 0.17 & 1.53 $\pm$ 0.07 & $\mathrm{1.551}^\mathrm{+0.032}_\mathrm{-0.036}$ & 1.40 $\pm$ 0.12\tnote{f}\\
        6 Lyn & -- & -- & -- &  $\mathrm{1.428}^\mathrm{+0.036}_\mathrm{-0.027}$ & 1.37 $\pm$ 0.14\tnote{f}\\
        HD 210702 &  1.71 $\pm$ 0.06 & 1.63 $\pm$ 0.13 & 1.58 $\pm$ 0.07& $\mathrm{1.604}^\mathrm{+0.038}_\mathrm{-0.034}$ & 1.47 $\pm$ 0.14\tnote{f}\\
        HD 4313 & 1.53 $\pm$ 0.09 & 1.35 $\pm$ 0.11 & 1.71 $\pm$ 0.13& $\mathrm{1.373}^\mathrm{+0.234}_\mathrm{-0.076}$& $\mathrm{1.61}^\mathrm{+0.13}_\mathrm{-0.12}$\tnote{g}\\
        HD 5319 & 1.28 $\pm$ 0.1 & 1.24 $\pm$ 0.14& -- & $\mathrm{1.278}^\mathrm{+0.089}_\mathrm{-0.149}$& $\mathrm{1.25}^\mathrm{+0.11}_\mathrm{-0.10}$\tnote{g}\\
        HD 106270 & 1.33 $\pm$ 0.05 & 1.33 $\pm$ 0.06 & --& $\mathrm{1.377}^\mathrm{+0.038}_\mathrm{-0.037}$ & $\mathrm{1.52}^\mathrm{+0.04}_\mathrm{-0.05}$\tnote{g}\\
        HD 145428 & -- & -- & -- & $\mathrm{0.930}^\mathrm{+0.076}_\mathrm{-0.022}$& $\mathrm{0.99}^\mathrm{+0.10}_\mathrm{-0.07}$\\
        HD 181342 & 1.7 $\pm$ 0.09 & 1.49 $\pm$ 0.19 & 1.78 $\pm$ 0.11 & 1.380 $\pm$ 0.120& $\mathrm{1.73}^\mathrm{+0.18}_\mathrm{-0.13}$\tnote{g}\\
        HD 185351 & -- & -- & 1.82 $\pm$ 0.05 & $\mathrm{1.687}^\mathrm{+0.043}_\mathrm{-0.221}$& $\mathrm{1.77}^\mathrm{+0.08}_\mathrm{-0.08}$\tnote{g}\\
        HD 212771 & 1.51 $\pm$ 0.08 & 1.22 $\pm$ 0.08 & 1.6 $\pm$ 0.13& $\mathrm{1.601}^\mathrm{+0.127}_\mathrm{-0.247}$ & $\mathrm{1.46}^\mathrm{+0.09}_\mathrm{-0.09}$\tnote{g}\\
		\bottomrule
	\end{tabularx}
	\begin{tablenotes}
	     \item[a]{Line list by \citet[TS13]{tsantaki_linelist_2013} are used for stars cooler than 5200 K while \citet[SO08]{sousa_linelist_2008} line list is used for hotter stars.}
	     \item[b]{Line lists by \citet[HM07]{hekker_linelist_2007} are used.}
	     \item[c]{Iron line lists by \citet[DS11]{dasilva_linelist_2011} are used.} 
	     \item[d]{\citet{stock_evoltime_2018} did not use a line list directly. Instead, they used the \teff, [Fe/H] and \logg\ values by \citet{hekker_linelist_2007}.}
	     \item[e]{{This work (also listed in Table~\ref{tab:data}).}}
	     \item[f]{Updated values of seismic masses from \citet{stello_asteroseismic_2017} (also listed in Table~\ref{tab:stello}). See Sec.~\ref{numax_err} for details.}
	     \item[g]{\citet{north_masses_2017}.}
	\end{tablenotes}
	\end{threeparttable}
\end{table*}

%%%%%%%%%%%%%%%%%%%%%%%%%%%%%%%%%%%%%%%%%%%%%%%%%%

% Don't change these lines
\bsp	% typesetting comment
\label{lastpage}
\end{document}